\documentclass[aps,prl,reprint,amsfonts,amssymb,amsmath]{revtex4-1}
\usepackage[hidelinks]{hyperref}
\usepackage{latexsym}
\usepackage[pdftex]{graphicx}

\newcommand\cO{{\cal O}}

\newcommand{\bea}{\begin{eqnarray}}
\newcommand{\eea}{\end{eqnarray}}

\newcommand{\psq}{\bar \Phi \Phi}

\begin{document}

\title{Bootstrapping the Three-Dimensional Supersymmetric Ising Model}
\date{\today}

\author{Nikolay Bobev${}^{1}$}
\author{Sheer El-Showk${}^{2,3,4}$}
\author{Dalimil Maz\'a\v c${}^{5}$}
\author{Miguel F. Paulos${}^{2}$}
\affiliation{${}^{1}$Instituut voor Theoretische Fysica, KU Leuven, Celestijnenlaan 200D, B-3001 Leuven, Belgium}
\affiliation{${}^{2}$Theory Division, CERN, Geneva, Switzerland}
\affiliation{${}^{3}$Sorbonne Universit\'{e}s, UPMC Univ Paris 06, UMR 7589 LPTHE, F-75005, Paris, France}
\affiliation{${}^{4}$CNRS, UMR 7589, LPTHE, 75005, Paris, France}
\affiliation{${}^{5}$Perimeter Institute for Theoretical Physics, Waterloo, Canada}

\begin{abstract}
\noindent We implement the conformal bootstrap program for three-dimensional CFTs with $\mathcal{N}=2$ supersymmetry and find universal constraints on the spectrum of operator dimensions in these theories. By studying the bounds on the dimension of the first scalar appearing in the OPE of a chiral and an anti-chiral primary, we find a kink at the expected location of the critical three-dimensional $\mathcal{N}=2$ Wess-Zumino model, which can be thought of as a supersymmetric analog of the critical Ising model. Focusing on this kink, we determine, to high accuracy, the low-lying spectrum of operator dimensions of the theory, as well as the stress-tensor two-point function. We find that the latter is in an excellent agreement with an exact computation.
\end{abstract}

\pacs{11.10.Kk, 11.30.-j, 11.30.Pb, 11.25.Tq}
\keywords{Two-dimensional conformal field theory, supersymmetry, R-symmetry, supergravity solutions}

\maketitle

\textit{Introduction.}--- Conformal field theories with $\mathcal{N}=2$ supersymmetry in 3d are interesting theoretical models with rich dynamics. For example, they enjoy a plethora of dualities akin to mirror symmetry in two dimensions and Seiberg duality in four dimensions \cite{deBoer:1997ka,deBoer:1997kr,Aharony:1997gp,Aharony:1997bx}. Via the AdS/CFT correspondence, they provide a window into the nonperturbative structure of M-theory, see for example \cite{Jafferis:2011zi}. These theories have also found applications in some areas of condensed matter physics, such as topological phases of matter \cite{Grover:2013rc} and optical lattices \cite{Yu:2010zv}. 

Of particular interest to us will be the Wess-Zumino (WZ) model with $\mathcal{N}=2$ supersymmetry. This is a theory of a complex scalar $\phi$, and a complex Dirac fermion $\psi$, with the Lagrangian 
\begin{equation}\label{WZaction}
\begin{split}
\mathcal{L}_{\text{WZ}} = \partial_{\mu}\bar{\phi} \partial^{\mu}\phi +i \bar{\psi}\gamma^{\mu}&\partial_{\mu}\psi + |\lambda|^2|\phi|^4 \\&+ (\lambda\phi\psi_{\alpha}\epsilon^{\alpha\beta}\psi_{\beta} + c.c.)\,.
\end{split}
\end{equation}
In superspace language, this is the theory of a single chiral superfield $\Upsilon=\phi+\theta \psi+\ldots$ with superpotential $\mathcal{W}=\Upsilon^3$. Supersymmetry is not spontaneously broken since the Witten index of this theory does not vanish. The coupling $\lambda$ is relevant  and the theory is believed to flow to a 3d $\mathcal{N}=2$ superconformal field theory (SCFT) in the infrared (IR), which we will denote by cWZ, for critical WZ model (see \cite{Strassler:2003qg} for a review). Note that the fermion cannot get a mass term because of the U(1) R-symmetry. Supersymmetry then guarantees the scalar stays massless as well. From \eqref{WZaction}, it is clear that this is a supersymmetric version of the critical Ising model, whose Lagrangian differs only by the absence of fermionic terms. This SCFT has been argued to arise as the IR fixed point of a certain lattice model \cite{Lee:2006if}, and also to describe a quantum critical point on the surface of a topological insulator \cite{Ponte:2012ru}. 

In this letter, we will use the techniques of the conformal bootstrap to find constraints on the space of $\mathcal{N}=2$ SCFTs in 3d. These methods have already been applied to theories with various amounts of supersymmetry in four dimensions \cite{Poland:2010wg,Vichi:2011ux,Poland:2011ey,Beem:2013qxa,Beem:2013sza,Berkooz:2014yda}, as well as for $\mathcal{N}=1$ \cite{Bashkirov:2013vya} and $\mathcal{N}=8$ \cite{Chester:2014fya,Chester:2014mea} theories in three dimensions. The analysis here is a natural continuation of our work in \cite{BEMP} which gives a broad-eyed view of SCFTs with four supercharges for general spacetime dimension $2\leq d<4$. In particular, in \cite{BEMP} we show that certain bounds on dimensions of operators show kinks, and we argue that one such kink should describe the cWZ model. Here we perform a detailed study of this kink for the dimension of phenomenological interest, $d=3$, following a similar analysis for the non-supersymmetric Ising model \cite{ElShowk:2012hu,El-Showk2014a}. The outcome is a precise evaluation of the spectrum of low-lying operators in the cWZ theory.
%
%\section{Bootstrap preliminaries}
%

\textit{Bootstrap preliminaries.}---We are interested in analyzing the consequences of crossing symmetry for a four-point function of the form $\langle \Phi\bar{\Phi} \Phi\bar{\Phi}\rangle$, where the operator $\Phi$ is a superconformal chiral primary operator, and $\bar{\Phi}$ is its conjugate. In other words, $\Phi$ is the lowest component of a short superconformal multiplet and its dimension is equal to its R-charge, $\Delta_{\Phi} = q_{\Phi}$. For the concrete example of the $\mathcal{N}=2$ WZ model in \eqref{WZaction}, $\Phi$ will be identified with the scalar field $\phi$ at the IR fixed point. 

The constraints from crossing symmetry are analysed in detail in \cite{BEMP}. Here we only outline the most salient features.
Decomposing the four-point function by performing an OPE expansion in the three inequivalent channels of fusing $\Phi$ with $\Phi$ and $\bar{\Phi}$ leads to two crossing equations.  From the $\Phi \times \bar \Phi$ OPE one finds that the four-point function decomposes into a sum of superconformal blocks,  $\mathcal{G}_{\Delta}^{s}$, corresponding to an exchanged superconformal primary of dimension $\Delta$ and spin $s$, and its superconformal descendants. The blocks take the form
\begin{equation}\label{eqn:scb}
\begin{aligned}
 \mathcal{G}_{\Delta}^{s}&(u,v) = G_{\Delta}^{s}(u,v)-\frac{\Delta+s}{2(\Delta+s+1)}G_{\Delta+1}^{s+1}(u,v) \\
 &-\frac{s^2(\Delta-s-1)}{2(4s^2-1)(\Delta-s)}G_{\Delta+1}^{s-1} (u,v) \\
 &+\frac{\Delta^2(\Delta+s)(\Delta-s-1)}{4(4\Delta^2-1)(\Delta+s+1)(\Delta-s)}G_{\Delta+2}^{s} (u,v)\,,
\end{aligned}
\end{equation}
%
%with the normalization condition $g_{\Delta}^s(u,v)\simeq \frac{(-1)^s}2\, u^{\frac{\Delta-s}2} (1-v)^s$ in the Euclidean OPE limit $x_1\to x_2$. 
where $G_{\Delta}^{s}(u,v)$ is the usual non-supersymmetric conformal block as a function of the two conformal invariant cross ratios, $u=\frac{x_{12}^2x_{34}^2}{x_{13}^2x_{24}^2}$ and $v=\frac{x_{13}^2x_{24}^2}{x_{13}^2x_{24}^2}$ \cite{Dolan:2000ut}. The representation in terms of ordinary blocks corresponds to the four allowed conformal primary operators obtained by acting with the supercharges.  

Due to the fact that $\Phi$ is chiral and using R-charge conservation, in the $\Phi \times \Phi$ OPE, we get a single operator per superconformal multiplet, and hence the four-point function decomposes as a sum of ordinary conformal blocks, $G_{\Delta}^s(u,v)$. The chirality condition also imposes that, in this channel, the OPE is of the schematic form (omitting conformal descendants)
\bea\label{phiphiope}
\Phi \times \Phi \simeq \Phi^2\,+Q^2 \bar \Psi\, +\ldots\;,
\eea
where $Q$ is one of the Poincar\'e supercharges, $\bar \Psi$ is an antichiral field with dimension $2(1-\Delta_\Phi)$, and $\Phi^2$ is chiral with dimension $2\Delta_{\Phi}$.  For the range of values of $\Delta_\Phi$ considered in this note, supersymmetry imposes a gap between the dimensions of these operators and the higher dimension or spin contributions denoted by `\ldots' in \eqref{phiphiope}.  Therefore, when checking crossing symmetry, we should allow for operators at these precise dimensions. Since we do not know their OPE coefficients, we can only specify that the operators {\em may} be present, not that they must be.  See \cite{BEMP} for more details.
\begin{figure}
\includegraphics[width=9 cm]{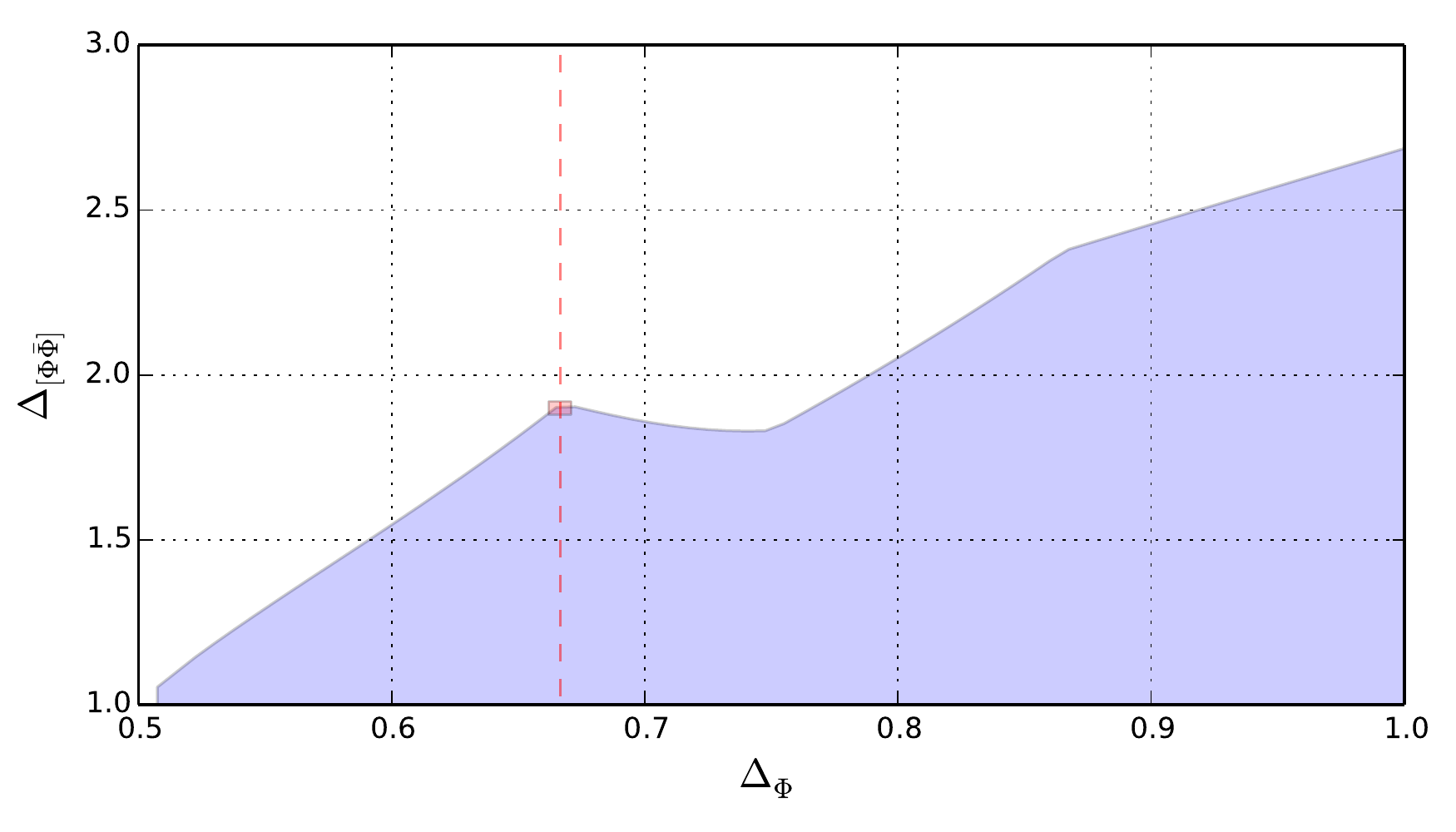}%
\caption{Bound on the dimensions of the leading unprotected operator in the $\Phi\times \bar \Phi$ OPE at $n_{max}=9$. There can be no unitary SCFTs in the white region. $\Delta_{\Phi}=2/3$ is indicated with a red dashed line.  The small shaded rectangle at $\Delta_\Phi=2/3$ indicates the field-of-view in Fig.\,\ref{fig:closeup}.}
\label{fig:generalbound}
\end{figure}
\begin{figure}
\includegraphics[width=9 cm]{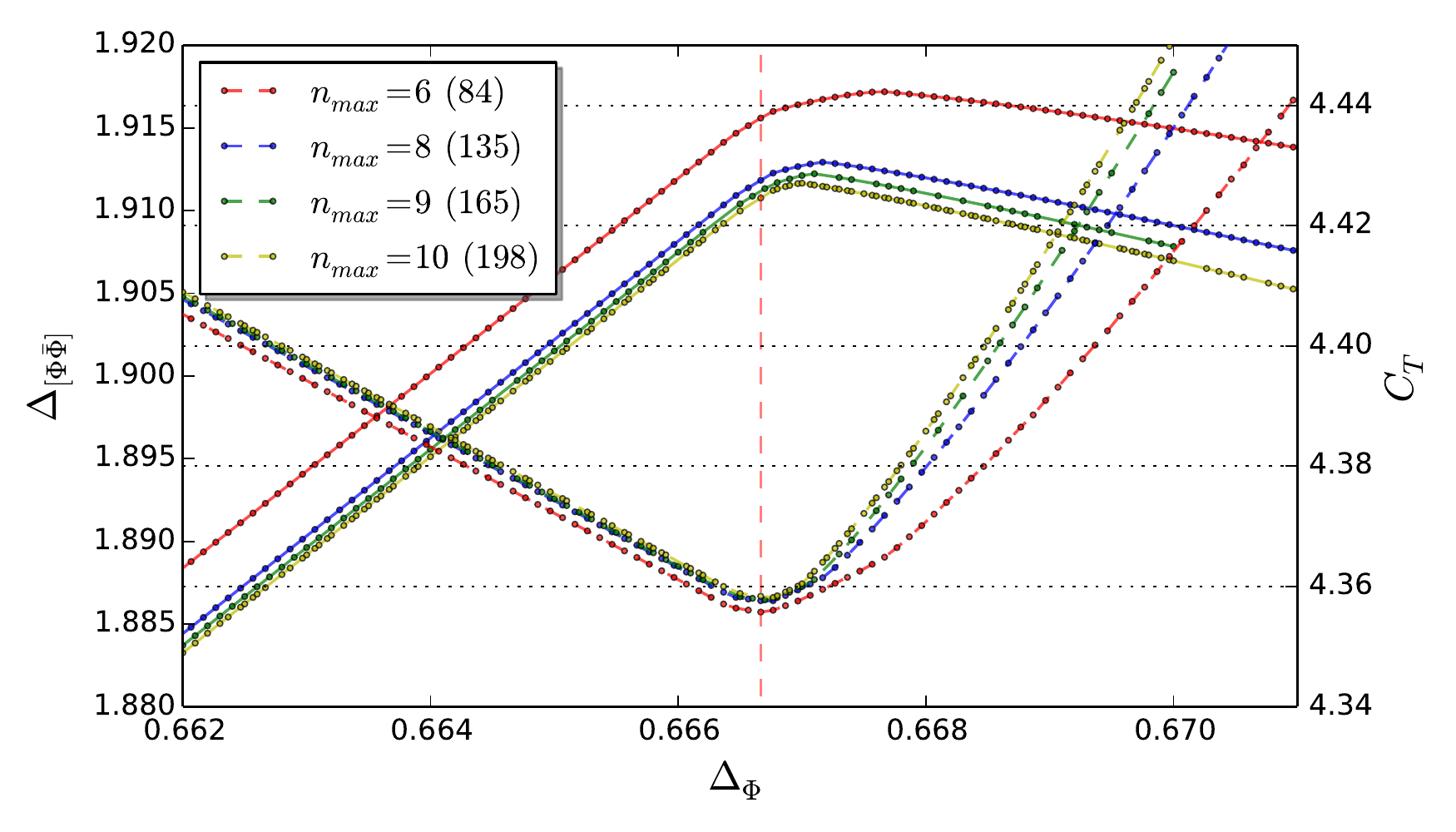}%
\caption{ Bound on the dimension of the leading unprotected operator in the $\Phi\times \bar \Phi$ OPE close to $\Delta_\Phi=2/3$ {\em (upper curves)}, and the corresponding central charge $C_T$ of the solution on the boundary {\em (lower curves)}. The numbers in parenthesis next to the values of $n_{max}$ indicate the number of constraints imposed to generate the associated curve.}
\label{fig:closeup}
\end{figure}
\begin{figure}
\includegraphics[width=9 cm]{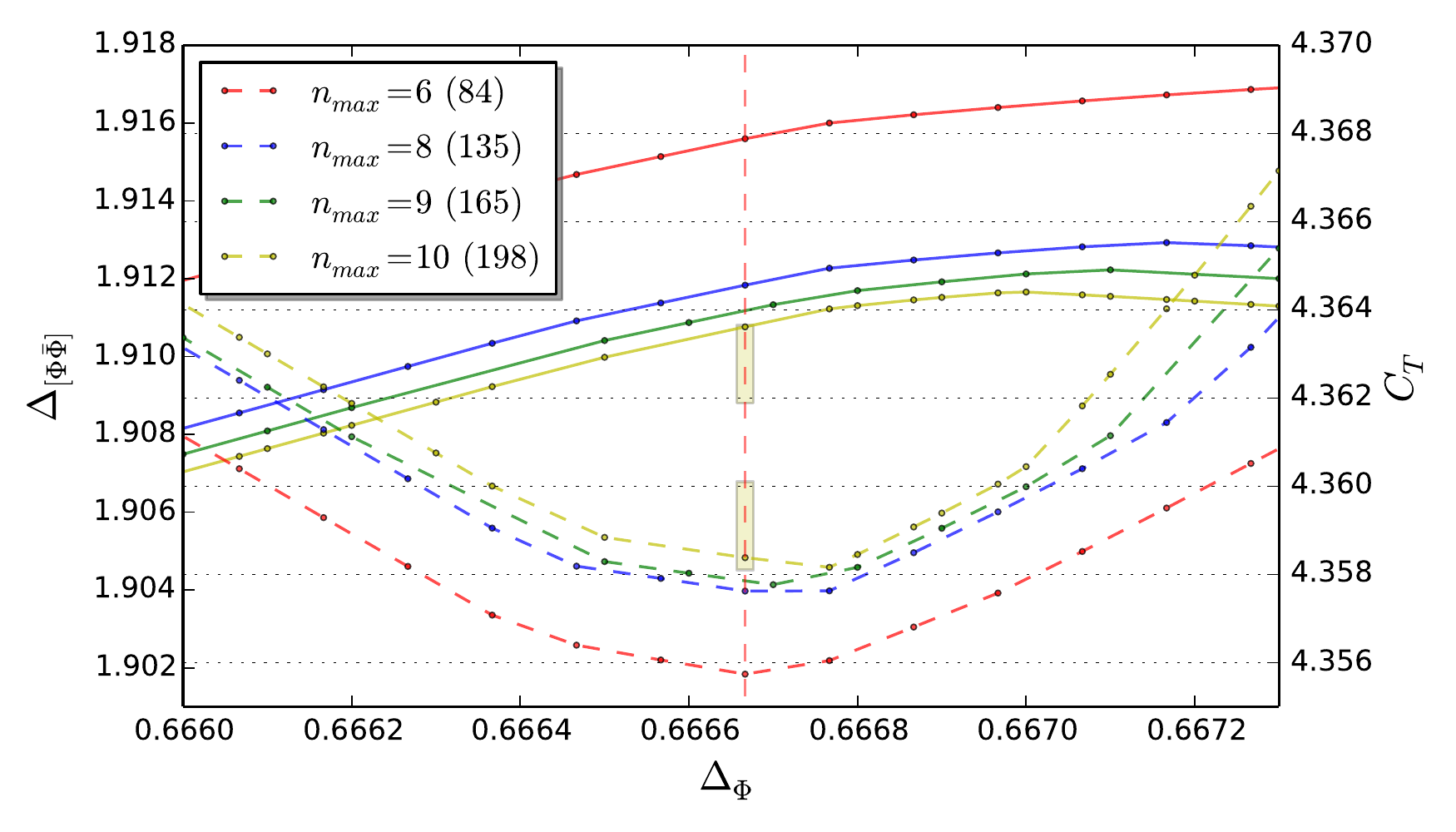}%
\caption{A closer view of Fig.\,\ref{fig:closeup}. The shaded rectangles indicate our estimated error for $\Delta_{[\Phi\bar{\Phi}]}$ and $C_T$.}
\label{fig:verycloseup}
\end{figure}

The two crossing equations can be written as an infinite number of linear equations with positive coefficients. The bootstrap analysis consists of checking under which conditions these equations may have a solution, essentially by solving a large linear program. This is done using a variation of Dantzig's simplex method implemented in Python \cite{El-Showk2014a}. Although the full set of constraints is continuously infinite, we can truncate the problem to a finite, discrete  subset obtained by considering finite Taylor expansions in the cross-ratios $u,v$ around some fixed point.  We parametrize the truncation by the integer $n_{max}$, with the actual number of constraints (terms in the Taylor expansion) growing essentially quadratically in this parameter (and indicated in parenthesis next to the value of $n_{max}$ in our figures). The reader is referred to \cite{BEMP,El-Showk2014a} for more details.

In the following we consider the problem of determining an upper bound on the dimension of the leading scalar superconformal primary in the $\Phi\times \bar\Phi$ OPE, which we denote by $[\Phi\bar\Phi]$. Notice that supersymmetry does not prevent this operator from having a large anomalous dimension. To derive such a bound we impose an ever-increasing a gap on the dimension of such operators in the OPE until crossing symmetry can no longer be satisfied. We only consider a finite number of constraints, but since each constraint is a {\em necessary} condition for crossing symmetry, this yields a rigorous (but potentially sub-optimal) upper bound on $\Delta_{[\Phi\bar\Phi]}$.  An important point is that if we tune the gap so that we sit precisely at the boundary of the allowed region (i.e. maximize the gap), it is possible to find a unique solution to the truncated constraint equations \cite{ElShowk:2012hu}. This yields a subset of the spectrum of operators and OPE coefficients.  As we increase the number of constraints this data changes as more operators and OPE coefficients are captured in a convergent manner \cite{ElShowk:2012hu}.  Below we argue that our bounds are saturated by cWZ, and are able to determine its low-lying spectrum with great accuracy.

%\begin{figure*}[hb]
\begin{figure}
\includegraphics[width=9 cm]{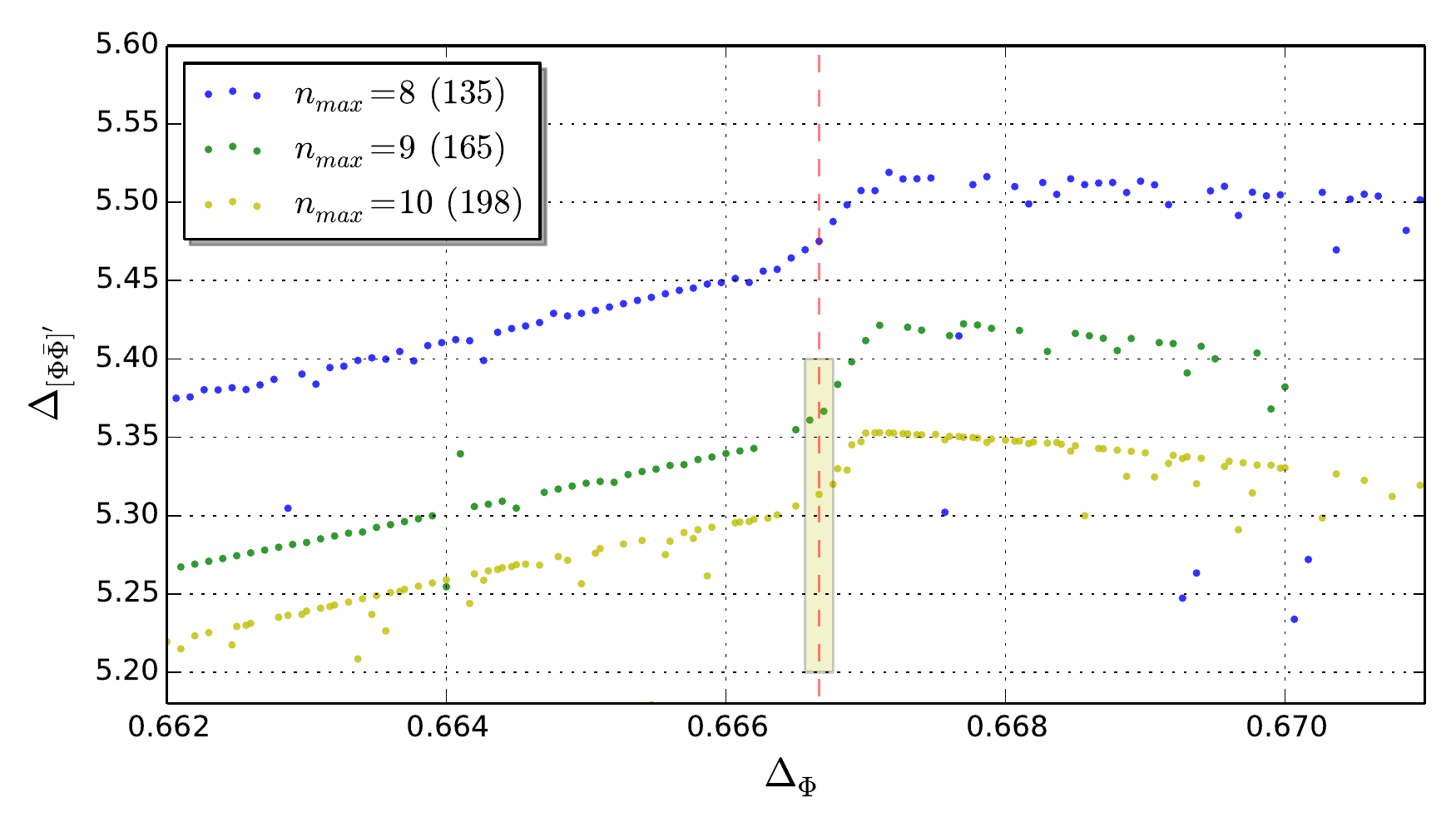}%
\caption{\label{fig:uncharged}Bound on the dimension of the subleading superconformal scalar primary, $[\Phi\bar\Phi]'$, in the $\Phi\times \bar \Phi$ OPE. The dimension is extracted from the solution that maximizes $\Delta_{[\Phi\bar\Phi]}$.}
\end{figure}
{\it Results.}--- In Fig.\,\ref{fig:generalbound} we show the results from the bootstrap analysis which lead to an upper bound on $\Delta_{[\Phi\bar{\Phi}]}$. There are three distinct features, which we refer to as kinks, in the curve bounding the allowed region. The second and third kink are discussed in more detail in \cite{BEMP}. Here we focus on the first kink, since it bears strong resemblance to a similar kink associated with the 3d critical Ising model \cite{El-Showk2014a}, and present the numerical results pertaining to the small box in Fig.\,\ref{fig:generalbound}.  

A close-up of this region is shown in Fig.\,\ref{fig:closeup} : the set of curves exhibiting an upward peak are the upper bounds on $\Delta_{[\Phi\bar{\Phi}]}$ for increasing values of $n_{max}$ (corresponding to imposing successively more bootstrap constraints).  Quite strikingly, we see that there is a sharp kink for $\Delta_\Phi\simeq 2/3$. This is precisely the (protected) dimension of the chiral field $\Phi$ in cWZ, since the superpotential $\mathcal{W}= \Upsilon^3$ fixes $\Delta_\Phi=2/3$. Hence it is natural to conjecture that our bound is actually saturated by cWZ at this point. As we increase the number of constraints imposed, the bounds converge rapidly towards this point. This is in line with previous bootstrap results that indicate that kinks in bounds correspond to actual CFTs.

In Fig.\,\ref{fig:closeup} we also display the central charge \footnote{Since there are no conformal anomalies in 3d we define the central charge as the coefficient of the two point function of the energy-momentum tensor.}, extracted from the OPE coefficients in the unique solution along the conformal dimension bound curve.  Here  we observe a sharp minimum which occurs very close to $\Delta_\Phi=2/3$. Such minima in the central charge have been observed before in the context of the non-supersymmetric bootstrap, where they were conjectured to signal the existence of the ordinary Ising model. Encountering the same phenomenon here provides further evidence for our conjecture that the values at the kink correspond to the cWZ theory. However, it is important to note that the plot in Fig.\,\ref{fig:closeup} does not correspond to central charge bounds as found in e.g. \cite{El-Showk2014a}, but rather indicates the value of the central charge for the unique solution that maximizes $\Delta_{[\Phi\bar{\Phi}]}$.

When examined at high resolution (Fig.\,\ref{fig:verycloseup})  we see that the horizontal positions of the two kinks do not exactly coincide, and they do not occur precisely at $\Delta_\Phi=2/3$. This is not inconsistent as we have only imposed a finite number of constraints (in parenthesis in the plot legends next to the value of $n_{max}$). It is clear from Fig.\,\ref{fig:closeup} that the location of the kinks is changing as a function of $n_{max}$ and seems to be converging towards $2/3$. As this note is an initial foray into the cWZ theory, we take a conservative approach and estimate the conformal dimensions from our most rigorous bounds at $\Delta_\Phi=2/3$ and then heuristically estimate the error from the rate of convergence.

\begin{figure}[hb]
\includegraphics[width=9 cm]{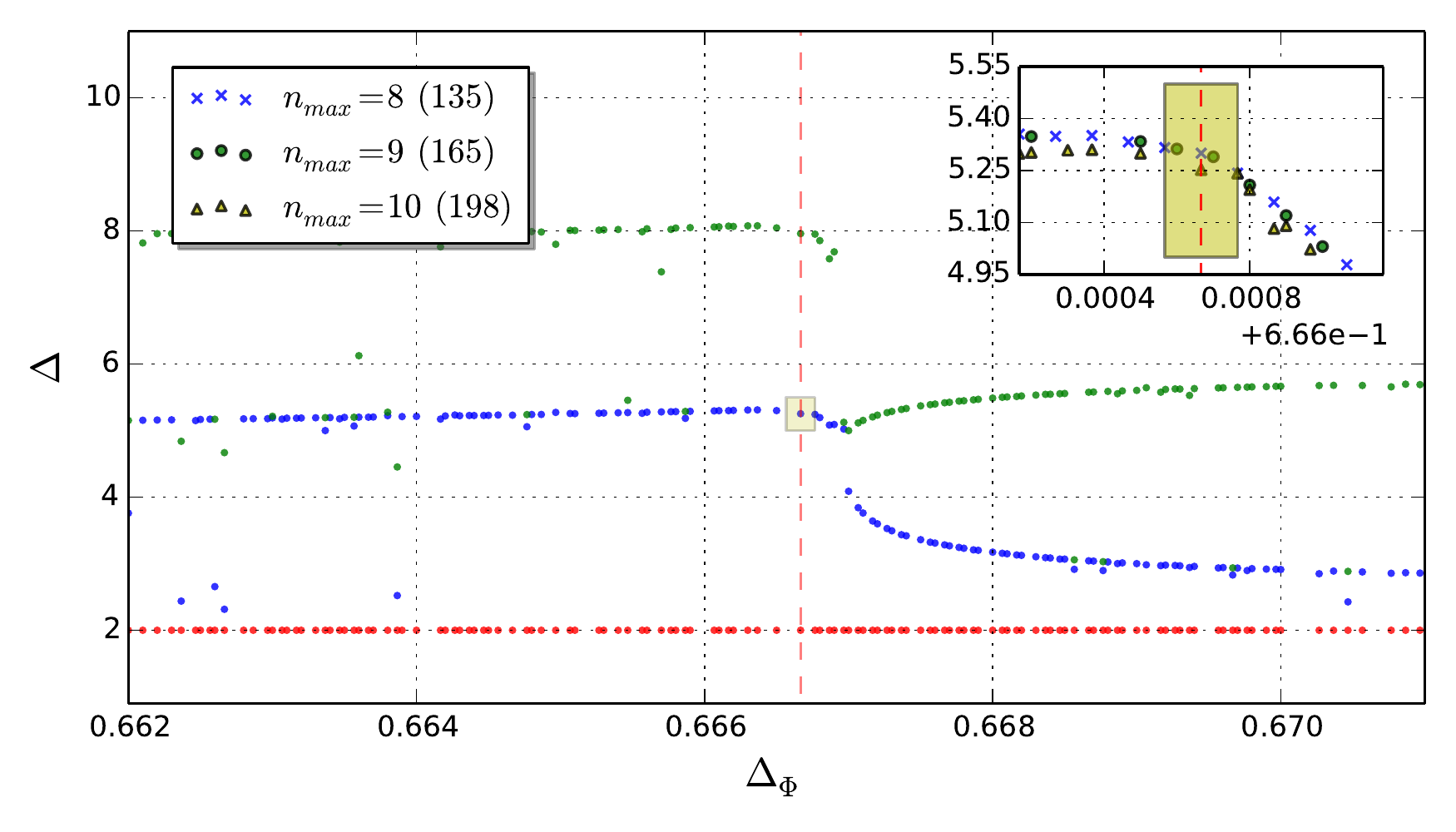}%
\caption{\label{fig:spin1}The spin-1 spectrum for $n_{max}=10$ in the vicinity of $\Delta_\Phi=2/3$.  The decoupling of the subleading spin-1 field is reminiscent of what happens in the spin-2 sector of the non-supersymmetric Ising model \cite{El-Showk2014a}.  In the inset we show the spin-1 spectrum for various values of $n_{max}$ which all lie well within our estimated error bars.}
\end{figure}

Our results for the low-lying spectrum are compiled in Table~\ref{tab:spec}, where we also present the relation between conformal dimensions and critical exponents.  To support our error estimates, we provide, in Fig.\,\ref{fig:verycloseup}, \ref{fig:uncharged}, and \ref{fig:spin1}, plots depicting the convergence rate of these different quantities.  Note that in the case of $\Delta_{[\Phi\bar{\Phi}]}$ our analysis yields a rigorous upper bound, since we know $\Delta_\Phi=2/3$ exactly; hence the error bars necessarily lie below our bound curve. We believe the analogous result is true for the central charge $C_T$. 
In Table~\ref{tab:spec}, we have also included the dimension of the non-superconformal primary operator $[Q^4\psq]$, obtained by acting with the four $Q$ supercharges on $[\psq]$.  This operator is usually called $\varepsilon'$ in the non-supersymmetric Ising model. Supersymmetry implies the relation $\Delta_{[Q^4\psq]} = \Delta_{[\psq]}+2$. Finally, $[\bar{\Phi}\Phi]'$ denotes the second-lowest nontrivial scalar superconformal primary and $J'$ the second-lowest spin-1 superconformal primary appearing in the OPE of $\Phi$ and $\bar{\Phi}$, the lowest being the R-current. Since $\Phi$, $[\bar{\Phi}\Phi]$, $[\bar{\Phi}\Phi]'$ and $J'$ are {\em superconformal} primaries, Table~\ref{tab:spec} implicitly yields the dimensions of their descendant conformal primaries.

\begin{table}
\caption{\label{tab:spec}Low-lying spectrum of the critical WZ model.}
\begin{ruledtabular}
\begin{tabular}{ll}
$\Delta_\Phi=1/2+\eta/2$ & $\frac 23$ (exact)\\
$\Delta_{[\psq]}=3-1/\nu $& $1.9098(20)$\\
$\Delta_{[Q^4\psq]} = 3+\omega $& $3.9098(20)$\\
$\Delta_{[\psq]'} $& $5.3(1)$\\
$\Delta_{J'} $& $5.25(25)$\\
$C_T $& $4.3591(20)$\\
\end{tabular}
\end{ruledtabular}
\end{table}
\begin{figure*}[ht]
\includegraphics[width=19 cm]{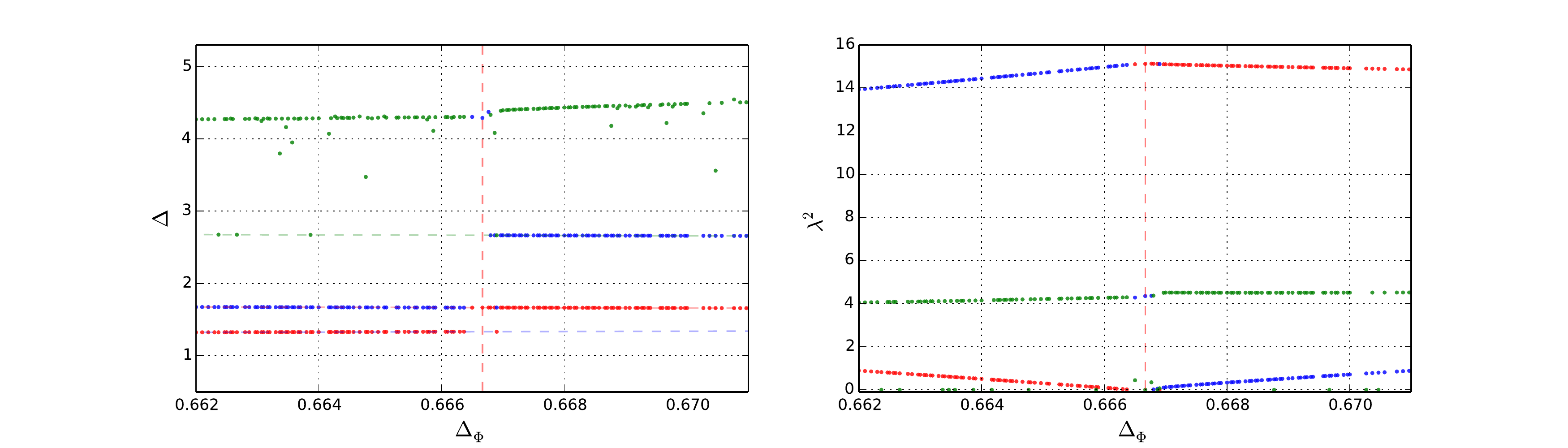}%
\caption{\label{fig:charged}Charged scalar spectrum in the vicinity of $\Delta_\Phi=2/3$.  {\em (Left:)} the first three spin zero operators in the spectrum (colored by order of appearance). The dashed lines correspond to $2\Delta_\Phi$, $d - 2 \Delta_\Phi$ and $2(d-1)-2\Delta_\Phi$ with $d=3$ (see \cite{BEMP}).  {\em (Right:)} the OPE coefficients for each operator appearing on the left hand plot (with matching colors).  Observe the vanishing of the $\Phi^2$ OPE coefficient at $\Delta_\Phi \simeq 2/3$.  The ``noisy'' operators in the spectrum plots can be seen to have vanishingly small OPE coefficients and thus correspond to small numerical artefacts in the solution.}
\end{figure*}

{\it Checks.}--- Unfortunately, there are not many results in the literature for the unprotected spectrum of cWZ. The only available result is a 1-loop $\epsilon$-expansion calculation \cite{Thomas,Lee:2006if}, which leads to the following low-lying spectrum \footnote{Identifying $\Phi$, $[\bar \Phi \Phi]$ with $\sigma$, $\varepsilon$ in those references.}:
\begin{equation}\label{1loopeps}
\Delta_{\Phi}= \dfrac{3-\epsilon}{3}\;, ~~~ \Delta_{[\Phi \bar\Phi]}= 2 + \cO(\epsilon^2)\;.
\end{equation}
The expression for $\Delta_{\Phi}$ is 1-loop exact due to supersymmetry and after setting $\epsilon=1$, one finds the expected result $\Delta_{\Phi}=2/3$. The fact that $\Delta_{[\Phi\bar{\Phi}]}=2$ in both $d=2$ and $d=4$ (see \cite{BEMP}), together with the expression in \eqref{1loopeps}, suggests that $\Delta_{[\Phi\bar{\Phi}]}$ never strays too far from 2, and indeed this is  within $5\%$ of our numerical estimate. 

A strong evidence supporting the identification of the kink as the cWZ model comes from the value of the stress-tensor two-point function $C_T$. Dividing by the value for the free chiral multiplet $C_T^{\mbox{\tiny free}} = 6$, the bootstrap prediction is $C_T^{\mbox{\tiny cWZ}}/C_T^{\mbox{\tiny free}}\simeq 0.7265(3)$. The same quantity can be computed exactly using supersymmetric localization \cite{Imamura:2011wg,Closset:2012ru,Nishioka:2013gza}, with the result $C_T^{\mbox{\tiny cWZ}}/C_T^{\mbox{\tiny free}}\simeq 0.7268$, showing excellent agreement with bootstrap. The ratio $C_T^{\mbox{\tiny cWZ}}/C_T^{\mbox{\tiny free}}$ is much smaller than the analogous quantity in the non-supersymmetric critical Ising model \cite{El-Showk2014a}, $\widehat{C}_T/\widehat{C}_T^{\mbox{\tiny free}}\simeq 0.95$.

An important characteristic of cWZ is the decoupling of the chiral operator $\Phi^2$ from the spectrum. This is a consequence of the cubic superpotential, which implies that $\Phi^2$ vanishes in the chiral ring. Hence, if our conjecture is correct, the spectrum at the kink must not contain this operator. To this end, in Fig.\,\ref{fig:charged}, we examine the OPE in the $\Phi \times \Phi$ channel close to the cWZ kink. Notice that the OPE coefficient of $\Phi^2$ goes to zero precisely at the kink. This feature may be thought of as the ultimate reason for the existence of the kink, by forcing a spectrum rearrangement at this point. The decoupling of $\Phi^2$ may also provide an explanation for the peculiar see-saw behaviour of the bound in Fig.\,\ref{fig:generalbound}. The next operator in \eqref{phiphiope} is $Q^2\bar \Psi$ whose dimension decreases with $\Delta_\Phi$ (it is given by $\Delta_{Q^2\bar \Psi}=3-2\Delta_\Phi$). It may be that this ``drags'' $\Delta_{\psq}$ down, explaining why, to the right of the kink, the bound decreases as we increase $\Delta_\Phi$.
%, a phenomena that has never been observed in such bootstrap bounds. NPB:I removed this comment since in the recent Beem et al 4d N=2 paper they see something like our plot but without the 3rd kink.
%
The decoupling of $\Phi^2$ implies also that the first operator in the $\Phi \times \Phi$ OPE is $Q^2 \bar \Psi$, which for $\Delta_\Phi=2/3$ has dimension $5/3=1+\Delta_\Phi$. This suggests the identification $\Psi\equiv\Phi$, so that the OPE takes the form
\bea
\Phi\times \Phi=Q^2 \bar \Phi+\ldots\,.
\eea
This is consistent with cWZ which has a single scalar chiral primary operator, and provides further evidence for our conjecture.

In Fig.\,\ref{fig:spin1}, we examine the spectrum of spin-1 operators in the $\Phi \times \bar \Phi$ OPE. Besides the existence of a conserved current (with $\Delta=2$), we observe rapid operator rearrangements as the kink is approached. It is interesting to compare this with what happens in the non-supersymmetric Ising model \cite{El-Showk2014a,ElShowk:2012ht}. There, one of the prominent features is the decoupling of a spin-2 operator with dimension $\simeq 3.5$ when approaching the kink from the right. Here we see the supersymmetric analog of this transition, with a spin-1 superconformal primary with dimension $\Delta\simeq 2.9$ decoupling from the right.  In supersymmetric theories many spin-2 fields, including the stress-tensor, are components of spin-1 superconformal multiplets so perhaps it is not so surprising that we observe a decoupling in the spin-1 sector.  It would be interesting to check if this decoupling has a $d=2$ analog coming from a Virasoro null state (as is the case in the non-supersymmetric model).
%Indeed in $d=2$\ldots
%(which contains a spin-2 primary in its multiplet) 

{\it Conclusions.}--- In this note, we have initiated the study of the cWZ theory using conformal bootstrap methods. Our results hinge on the conjecture that this theory sits at the kink in our numerical bounds. Bearing this in mind, we have provided the most accurate calculation to date of the critical exponents and OPE coefficients in the cWZ model. It will certainly be desirable to corroborate our analysis by using other methods, like the $\epsilon$-expansion or Monte Carlo estimates.  It will also be interesting to apply the alternative conformal bootstrap algorithm of \cite{Gliozzi:2013ysa} to 3d $\mathcal{N}=2$ theories.

\begin{acknowledgements}
We would like to thank C. Beem, D. Gaiotto, S. Giombi, I. Klebanov, M. Meineri, S.-S. Lee, S. Pufu, S. Rychkov, B. van Rees and A. Vichi for useful and informative discussions. The work of NB is supported by a starting grant BOF/STG/14/032 from KU Leuven. The research of DM was supported by Perimeter Institute for Theoretical Physics. Research at Perimeter Institute is supported by the Government of Canada through Industry Canada and by the Province of Ontario through the Ministry of Research and Innovation.  During this work MFP was supported by a Marie Curie Intra-European Fellowship of the European
Community's 7th Framework Programme under contract number PIEF-GA-2013-623606 and by US DOE-grant DE-SC0010010. Numerical calculations were performed on the CERN cluster and Brown University's CCV cluster.
\end{acknowledgements}

\bibliography{3dPRL}
\end{document}